\input epsf.tex
\documentstyle[12pt]{article}
\begin{document}
\newcommand\bfigt{\begin{figure}[top]}
\newcommand\efig{\end{figure}}
\newcommand{\ra}{\rightarrow}
\newcommand{\tn}{\otimes}
\newcommand{\ga}{\gamma}
\newcommand{\pl}{\partial}
\newcommand{\la}{\lambda}
\newcommand{\bq}{\begin{equation}}
\newcommand{\eq}{\end{equation}}
\newcommand{\ph}{\varphi}
\newcommand{\iy}{\infty}
\newcommand{\dl}{\delta}
\def\up#1{\leavevmode \raise .3ex\hbox{$#1$}}
\def\down#1{\leavevmode \lower .4ex\hbox{$\scriptstyle#1$}}
\def\ch{\up{\chi}_{\down{J}}}
\newcommand{\ov}{\over}
\newcommand{\ve}{\varepsilon}
\newcommand{\De}{\Delta}
\newcommand{\rh}{\rho}
\newcommand{\ba}{\left(\begin{array}{cc}}
\newcommand{\ea}{\end{array}\right)}
\newcommand{\bdet}{\left|\begin{array}{cc}}
\newcommand{\edet}{\end{array}\right|}
\newcommand{\eph}{\varepsilon\varphi}
\newcommand{\eps}{\varepsilon\psi}
\newcommand{\be}{\beta}
\newcommand{\al}{\alpha}
\newcommand{\hf}{{1\ov2}}
\newcommand{\noi}{\noindent}
\newcommand{\lp}{\left(}
\newcommand{\rp}{\right)}
\begin{center}
{\large \bf The Distribution of the  Largest Eigenvalue \\
 in the Gaussian Ensembles:~\mbox{\bf $\beta=1,2,4$} }
 \end{center}
\begin{center}{{\bf Craig A. Tracy}\\
{\it Department of Mathematics and Institute of Theoretical Dynamics\\
University of California, Davis, CA 95616, USA\\
e-mail address: tracy@itd.ucdavis.edu}}\end{center}
\begin{center}{{\bf Harold Widom}\\
{\it Department of Mathematics\\
University of California, Santa Cruz, CA 95064, USA\\
e-mail address: widom@math.ucsc.edu}}\end{center}
\begin{abstract}
 The focus of this survey paper is on the  distribution
 function $F_{N\be}(t)$ for the largest eigenvalue in the
 finite $N$ Gaussian Orthogonal Ensemble (GOE, $\be=1$), the Gaussian
 Unitary Ensemble (GUE, $\be=2$), and the Gaussian Symplectic
 Ensemble (GSE, $\be=4$) in the edge scaling limit of $N\ra\iy$.
 These limiting distribution functions are expressible in
 terms of a particular Painlev{\'e} II function.  Comparisons
 are made with finite $N$ simulations as well as a discussion of
 the universality of these distribution functions.
\end{abstract}
\par\par
\renewcommand{\theequation}{1.\arabic{equation}}
\noi{\bf 1. Introduction}\par
In the well known Gaussian random matrix models  of
Wigner-Dyson-Mehta \cite{mehta,porter,intro},
 the probability
density that the eigenvalues lie in infinitesimal intervals about
the points $x_1,\ldots,x_N$ is given by
\[P_{N\be}(x_1,\ldots,x_N)=C_{N\be}\; e^{-{1\over 2}\be\sum x_i^2}\,
\prod_{j<k}|x_j-x_k|^{\be},\]
where  $C_{N\be}$ is a normalization constant and
\[
\be:=\left\{ \begin{array}{lll} 1 \mbox{\rm\quad for GOE,} \\
                               2 \mbox{\rm\quad for GUE,  }\\
                               4 \mbox{\rm\quad  for GSE.}
             \end{array}
             \right. \]
We recall that for $\be=1$ the matrices are $N\times N$ real
symmetric, for $\be=2$ the matrices are $N\times N$ complex
Hermitian, and for $\be=4$ the matrices are $2N\times 2N$
self-dual Hermitian matrices. (For $\be=4$ each eigenvalue
has multiplicity two.)
\par
We are interested in
\begin{eqnarray}
 E_{N\be}(0;J)& := &\raisebox{-4mm}{$\displaystyle
 \int\cdots\int \atop \hspace{-2mm} x_j{\not\in} J$}
  P_{N\be}(x_1,\ldots,x_N)\, dx_1 \cdots dx_N \label{prob1}\\
 & = &\mbox{\rm probability no eigenvalues lie in $J$.}\nonumber
 \end{eqnarray}
 For $J=(t,\iy)$,
$F_{N\be}(t):=E_{N\be}(0;(t,\iy))$ is the
distribution function for the largest eigenvalue, that is,
\[ \mbox{Prob}(\lambda_{\rm max}<t)=F_{N\be}(t).\]
If we introduce the $n$-point correlations
\[
 R_{n\be}(x_1,\ldots,x_n):=
 {N!\ov (N-n)!}
\int_{\bf R}\cdots\int_{\bf R}
   P_{N\be}(x_1,\ldots,x_N)\, dx_{n+1} \cdots dx_N \, ,
\]
and  we denote by $\ch$ the characteristic function of the set $J$,
then we may rewrite (\ref{prob1}) as
\begin{eqnarray}
E_{N\be}(0;J)&=&
\int_{\bf R}\cdots\int_{\bf R}  P_{N\be}(x_1,\ldots,x_N) \prod_i (1-\ch(x_i))
\, dx_1\cdots dx_N \nonumber \\
&=& 1 - \int_J R_{1\be}(x_1)\, dx_1 +
 {1\ov 2!} \int_J\int_J R_{2\be}(x_1,x_2)\,dx_1 dx_2 -\vspace{10mm} \nonumber\\
&& {1\ov 3!} \int_J\int_J\int_J R_{3\be}(x_1,x_2,x_3) \, dx_1 dx_2 dx_3
+ \cdots \label{prob2}
\end{eqnarray}
\par
If we reinterpret $P_{N\be}$ as the equilibrium Gibbs measure for $N$
like charges interacting with a logarithmic Coulomb potential (confined to
the real line) subject to a harmonic confining potential,  then everything we
have said so
far is valid for arbitrary inverse temperature $\be>0$.  In this
interpretation $F_{N\be}(t)$ is the probability, at inverse
temperature $\be$, that the interval $(t,\iy)$ is free from charge.
\par
\setcounter{equation}{0}\renewcommand{\theequation}{2.\arabic{equation}}
\noi{\bf 2.
Why $\beta=2$ is the Simplest Case}
\par
For $\be=2$  the $n$-point functions  take a particularly
simple form \cite{mehta}
\bq R_{n2}(x_1,\ldots,x_n)=\det
\left(K_N(x_i,x_j\right)\vert_{i,j=1}^n)\label{R2}\eq
where
\begin{eqnarray*}
 K_N(x,y)&:=&\sum_{i=0}^{N-1}\,\ph_i(x)\,\ph_i(y) \\
 &=&{\ph(x) \psi(y) - \psi(x) \ph(y) \ov x-y },
 \end{eqnarray*}
  $\ph_i(x)$ are the orthonormal harmonic oscillator wave functions, i.e.\
 $\ph_k(x) := c_k e^{-{1\ov 2} x^2} H_k(x)$, $H_k$ Hermite polynomials, and
 \[\ph(x):=\left({N\ov2}\right)^{1/4}\,\ph_N(x),
 \quad\psi(x):=\left({N\ov2}\right)^{1/4}\,\ph_{N-1}(x).\]
 Using (\ref{R2}) in (\ref{prob2}) we see that this expansion
 is the Fredholm expansion
 of the operator $S$ whose kernel is $K_N(x,y)\ch(y)$; that is,
 \[ E_{N2}(0;J) = \det(I-S). \]
 \par
 At this point we can use the general  theory \cite{FD,systems}  for Fredholm
determinants
 of operators $K$
 whose kernel is of the form
\[ {\ph(x) \psi(y) - \psi(x) \ph(y) \ov x-y } \ch(y) \]
where the $\ph$ and $\psi$ are assumed to satisfy
\[ {d\ov dx} {\ph \choose \psi} = \Omega(x) {\ph \choose \psi} \]
with $\Omega(x)$  a $2\times 2$ matrix, trace zero, with  rational entries in
 $x$.
 It is shown in \cite{FD,systems} that if
 \[ J:=\bigcup_{j=1}^m \left(a_{2j-1},a_{2j}\right), \]
 then
 \[ {\pl\ov \pl a_j} \log\det\left(I-K\right),\> j=1,\ldots,2m, \]
 are expressible polynomially in terms of solutions to a total system
 of partial differential equations ($a_j$ are the independent
 variables).
 \par
 In the finite $N$ GUE case
\[ \Omega(x) = \left(\begin{array}{cc} -x & \sqrt{2N} \\
					-\sqrt{2N} & x \end{array}
					\right)\]
 and the theory gives for $J=(t,\iy)$
 \[ F_{N2}(t) = \exp\left(-\int_t^\iy R(x)\, dx\right) \]
 where $R$ satisfies ( $^\prime=d/dt$)
 \[(R^{\prime\prime})^2 + 4(R^{\prime})^2(R^{\prime}+2N)-4(t R^{\prime}-R)^2=0.
\]
This last differential equation is the
 $\sigma$ version of Painlev{\'e} IV ($P_{IV}$) \cite{jm,okamoto}.  What is
remarkable
 about this  result is that the size of the matrix, $N$, (equivalently
 the dimension of the integral (\ref{prob1})) enters only as a coefficient
 in the above second order equation.
 \par
\setcounter{equation}{0}\renewcommand{\theequation}{3.\arabic{equation}}
\noi{\bf 3. Edge Scaling Limit}
 \par
 The famous Wigner Semicircle Law states that if $\rho_N(x)$ is the density of
 eigenvalues in any of the three Gaussian ensembles, then
 \[ \lim_{N_\be\ra\iy} {1\ov 2 \sigma \sqrt{N_\be}}
\rho_N\left(2\sigma\sqrt{N_\be}\, x\right) =
 \left\{ \begin{array}{ll} {1\ov \pi}\sqrt{1-x^2} & \vert x \vert < 1, \\
 				0 & \vert x \vert > 1.
 				\end{array}\right. \]
 Here $\sigma$, $\sigma/\sqrt{2}$, $\sigma/\sqrt{2}$ (for $\be=1,2,4$,
respectively)
 is the standard deviation of the Gaussian distribution in the off-diagonal
 elements and
\[ N_\be=\left\{ \begin{array}{lr} N,\> \be=1, \\
                               N,\> \be=2,\\
                               2N+1,\>\be=4.
             \end{array}
             \right. \]
   For the normalization here and in \cite{mehta}, \[\sigma=1/\sqrt{2}.\]
 Perhaps less well known is that the following is also true \cite{BY}:

 \[ F_{N\be}\left(2\sigma\sqrt{N_\be}+x\right) \ra \left\{\begin{array}{ll} 0 &
\mbox{if $x<0$} \\
                                             1 & \mbox{if $x>0$}
                                             \end{array}\right.\]
  as $N\ra\iy$.
  The edge scaling variable, $s$,  gives the scale on which to study
fluctuations
  \cite{brezin,forrester,airy}
  \bq t= 2\sigma\sqrt{N_\be} + {\sigma\> s \ov N_\be^{1/6} }\>
.\label{sVariable}\eq
  The edge scaling limit is the limit $N\ra\iy$, $s$ fixed, and in this limit
\cite{airy,FD}
 \[ F_{N2}(t)\ra \exp\left(-\int_s^\iy (x-s) q(x)^2 \, dx\right) =: F_2(s)\]
  where $q$ is the solution to the $P_{II}$ equation
\bq q''=s\,q+2\,q^3\label{P2}\eq
 satisfying the condition
 \bq q(s)\sim {\rm Ai}(s) \ \ \mbox{as}\ \ s\ra\iy, \label{BC}\eq
 with $\mbox{Ai}$ the Airy function.
\par
 We note that $F_2(s)$ is  the $P_{II}\>$  $\tau$-function \cite{jm,okamoto}.
 There is a rather complete description of the
  one-parameter family of solutions to $P_{II}$  satisfying the condition
 \[q(s;\lambda)\sim \lambda {\rm Ai}(s)\] as $s\ra\iy$
 and an analysis of
 the corresponding connection problem for $s\ra -\iy$ (see
\cite{clarkson,deift} and
 references therein).
 \par
\setcounter{equation}{0}\renewcommand{\theequation}{4.\arabic{equation}}
\noi{\bf 4. Cases $\be$ = 1 and 4 }
 \par
 In \cite{dyson} Dyson showed  for the circular ensembles
 with $\be=1$ or $4$
 that the $n$-point correlations could be written as
 \[ \left(R_{n\be}(x_1,\ldots,x_n)\right)^2 =\det\left(K_{N\be}(x_i,x_j)
 \vert_{i,j=1}^n \right) \]
 where now
 \[ K_{N\be}(x,y) = 2\times 2\> \mbox{matrix.}\]
 \par
 Mehta  generalized this result to the finite $N$ Gaussian ensembles
\cite{mehta} and
 Mahoux and Mehta \cite{mahoux}
  gave a general method for invariant matrix models for both $\be=1$ and
 $\be=4$.
Mehta's result implies that for $\be=1$ or $4$ that
 \[ \left(E_{N\be}(0;J)\right)^2 = \det\left(I-K_{N\be}\right) \]
 where  $K_{N\be}$ is an operator with $2\times 2$ matrix kernel, or
equivalently a
 $2\times 2$ matrix with operator entries.
 Explicitly for $\be=1$ (and $N$ even):
 \[ K_{N1}=\ch\ba S+\psi\tn\ve\ph&SD-\psi\tn\ph\\&\\\ve S-\ve+\ve\psi\tn\ve\ph
&S+\ve\ph\tn\psi\ea\ch,\]
where $S$, $\ph$ and $\psi$ are as before for $\be=2$, and
\begin{eqnarray*}
\ve f(x)&:=&\int_{-\iy}^\iy \ve(x-y) f(y)\, dy \, , \\
Df(x)&:=& {df(x)\ov dx}\, ,
\end{eqnarray*}
with
\[ \ve(x):=\left\{\begin{array}{lll} \;{ 1\ov 2} & \mbox{if $x>0$,} \\
					\; 0 & \mbox{if $x=0$,} \\
					\hspace{-.95ex} -{1\ov 2} & \mbox{if $x<0$.}
			\end{array}\right.\]
We have used the notation $a\tn b$ for the operator with kernel $a(x)\,b(y)$.

For $\be=4$, $\left(E_4(0;J/\sqrt{2})\right)^2$
is again a Fredholm determinant with
\[ K_{N4}= {1\ov 2} \ch\ba S+\psi\tn\ve\ph&SD-\psi\tn\ph\\&\\\ve
S+\ve\psi\tn\ve\ph
&S+\ve\ph\tn\psi\ea\ch.\]
In \cite{orthog} these Fredholm determinants are related to integrable systems
for general $J$.  Although \cite{orthog} treats exclusively the Gaussian
ensembles,
the methods appear quite general and should apply to other ensembles as well.
\par
\setcounter{equation}{0}\renewcommand{\theequation}{5.\arabic{equation}}
\noi{\bf 5. Idea of Proof and Results}
\par
Our derivation \cite{orthog} rests on the fact that operator determinants may
be manipulated much as
scalar determinants, as long as one exercises some care. We first write
 \[
 K_{N1}=\ba \ch D&0\\&\\0&\ch\ea
\ba(S\ve-\eph\tn\eps)\ch&(S+\eph\tn\psi)\ch\\&\\(S\ve-\ve-\eph\tn\eps)\ch
&(S+\eph\tn\psi)\ch\ea.
\]
This uses the trivial fact that $D\ve=I$ and the commutator identities
\[[S,\,D]=\ph\tn\psi+\psi\tn\ph,\quad[\ve,\,S]=-\eph\tn\eps-\eps\tn\eph.\]
By the general identity $\det(I-AB)=\det(I-BA)$ the determinant is
unchanged if the
factors are interchanged and so we may work instead with
 \[\ba (S\ve-\eph\tn\eps)\ch D&(S+\eph\tn\psi)\ch\\&\\
(S\ve-\ve-\eph\tn\eps)\ch D&(S+\eph\tn\psi)\ch \ea .\]
Applying a pair of row and column operations, which is justified by the
 identity $\det(I-A)=\det(I-BAB^{-1})$,
 we reduce this to
 \[\ba(S\ve-\eph\tn\eps)\ch D+(S+\eph\tn\psi)\ch&(S+\eph\tn\psi)\ch\\&\\
-\ve\ch D&0 \ea .\]
More row and column operations, which this time use $\det(I-A)(I-B)=
\det(I-A)\det(I-B)$, followed by factoring out
\[ I- S \ch\]
show that $E_{N1}(0;J)^2$ equals $E_{N2}(0;J)$ times a determinant of the
general
form
\[\det\,(I-\sum_{k=1}^n\al_k\tn\be_k),\]
whose value equals that of the scalar determinant
\[\det\,\Bigl(\dl_{j,k}-(\al_j,\be_k)\Bigr)_
{j,k=1,\cdots,n}.\]
Thus the evaluation of $E_{N1}(0;J)$ is reduced to the evaluation of certain
inner products and the result for $E_{N2}(0;J)$.
\par
To evaluate the inner products differential equations are derived for them.
For $J=(t,\iy)$ and in the  edge scaling limit these
differential equations can be solved with the result that
\bq F_1(s)^2=F_2(s)\,e^{-\int_s^{\iy}q(x)\,dx}\label{F1}\eq
where we recall
\bq F_2(s)=\exp\left(-\int_s^{\iy}(x-s)\,q(x)^2\,dx\right)\> .\label{F2}\eq
Similary for $\be=4$
\bq\small F_4(s/\sqrt{2})^2=F_2(s)\,\lp{e^{\hf\int_s^{\iy}q(x)\,dx}+
e^{-\hf\int_s^{\iy}q(x)\,dx}
\ov2}\rp^2\> .\label{F4}\eq
We note that there are no adjustable parameters in $F_\be$.
\par
Using the known asymptotics  of $q(s)$ as $s\ra\pm\iy$,
it is straighforward to solve (\ref{P2}) numerically and
produce accurate numerical values for $F_\be(s)$ and
the corresponding densities $f_\be(s)=dF_\be/ds$.  The densities
for the largest eigenvalue in each of the three ensembles
in the edge scaling limit
are displayed in Fig.~1. Observe that for higher ``temperature''
(lower $\be$) the variance of $f_\be(s)$ increases as one
would expect from the Coulomb gas interpretation.
\bfigt\vspace*{-15mm}\hspace*{10mm}\epsfysize=80mm \epsffile{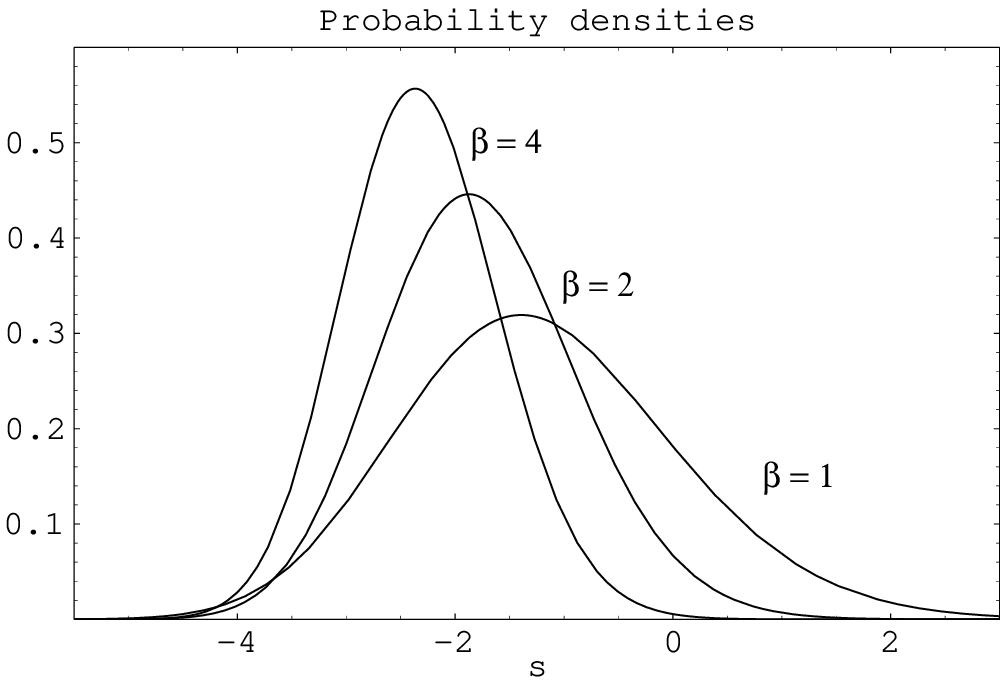}
\vspace*{-5mm}
\caption{The probability densities $f_\be(s)=dF_\be/ds$,
$\be=1,2,4$,  for the position of the largest eigenvalue
in the edge scaling limit.}
\efig
In Table 1 we give the mean ($\mu_\be$), standard deviation ($\sigma_\be$),
skewness ($S_\be$),
and the kurtosis ($K_\be$)\ of the
densities $f_\be$.\footnote[1]{The skewness of a density $f$ is $\int
\left({x-\mu\ov\sigma}
\right)^3 f(x)\,dx$ and the kurtosis is $\int\left({x-\mu\ov\sigma}\right)^4
f(x)\,dx - 3 $
where the $-3$ term makes the value zero for the normal distribution.}
\par
\setcounter{equation}{0}\renewcommand{\theequation}{6.\arabic{equation}}
\noi{\bf 6.
 How ``Useful'' are these  Scaling Functions?}
\par

Using a random number generator to simulate a  real symmetric
matrix of size $N\times N$ in the GOE,
we  calculate its largest eigenvalue
$\lambda_{\rm max}$.
For  $N$ large  the expected value  should be approximately
\[ E\left(\lambda_{\rm max}\right) = 2\sigma\sqrt{N} + {\sigma \mu_1\ov
N^{1/6}} \]
where $\mu_1$ is the mean of $f_1$ (see Table I).
In one such simulation (with $\sigma=1/\sqrt{2}$) with $N=500$ and 5000 trials
the mean
largest eigenvalue is $31.3062$ which should be compared with the theoretical
prediction of $31.353$.
\par
For each largest eigenvalue $\lambda_{\rm max}$ we compute a scaled
largest eigenvalue $s$ from
\[ \lambda_{\rm max} = 2\sigma\sqrt{N} + {\sigma s\ov N^{1/6}} \]
and form a histogram of $s$ values.  In Fig.~2 we compare this histogram
with the limiting density $f_1(s)$.  In Table II we give
the mean, standard deviation, skewness and kurtosis of this same scaled
largest eigenvalue data.  This should
be compared with the limiting theoretical values in the first row of Table I.
Similarly in Fig.~3 we
give comparisons of finite $N (=100)$ simulations in GUE and GSE with the
limiting
densities $f_2$ and $f_4$, respectively. The descriptive statistics
of this data are given in Table II.
\par
\begin{table}
\begin{center}
\begin{tabular}{|l|cccc|}\hline
$\be$ & $\mu_\be$ & $\sigma_\be$ & $S_\be$ & $K_\be$ \\ \hline\hline
1 & -1.20653 & 1.2680 & 0.293 & 0.165 \\ \hline
2 & -1.77109 & 0.9018 & 0.224 & 0.093 \\ \hline
4 & -2.30688 & 0.7195 & 0.166 & 0.050 \\ \hline
\end{tabular}
\caption{The mean ($\mu_\be$), the standard deviation ($\sigma_\be$), the
skewness ($S_\be$) and the kurtosis ($K_\be$) for the the densities $f_\be$.}
\end{center}
\end{table}
\par
\bfigt\vspace*{-10mm}\hspace*{10mm}\epsfysize=80mm \epsffile{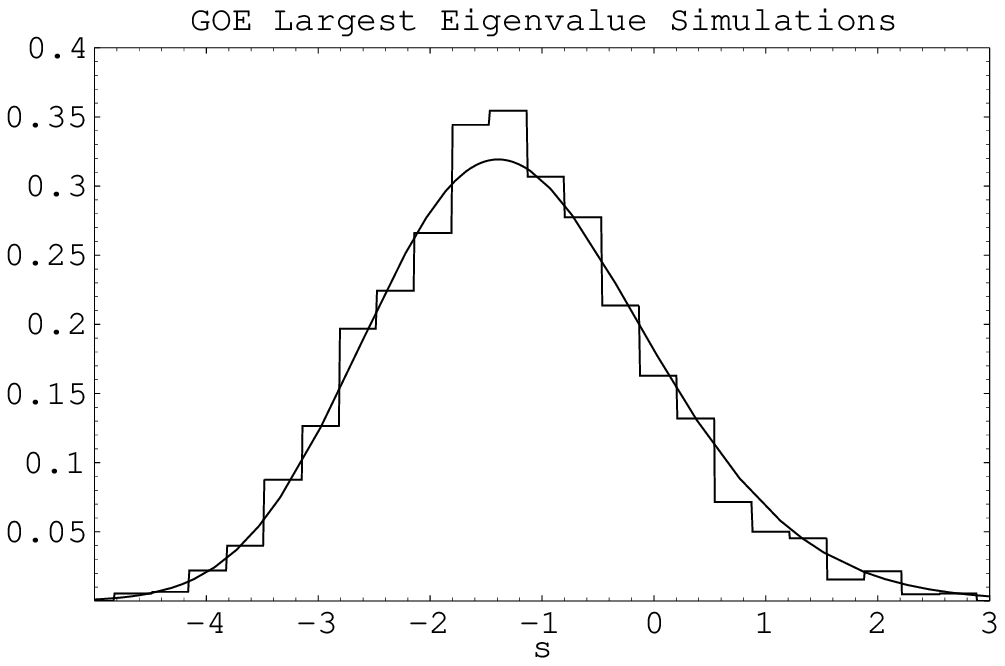}
\caption{Histogram of scaled largest eigenvalues for $N=500$, GOE for
5000 trials.  Solid
curve is the limiting density $f_1(s)$.}
\vspace*{-5mm}
\efig
\begin{table}
\vspace*{-25mm}
\begin{center}
\begin{tabular}{|l|cccc|}\hline
$\be$ & $\mu_\be$ & $\sigma_\be$ & $S_\be$ & $K_\be$ \\ \hline\hline
1 & -1.2614 & 1.2182  & 0.273 & 0.171 \\ \hline
2 & -1.7772 & 0.8952 & 0.133 & -0.089 \\ \hline
4 & -2.3154 & 0.7057 & 0.171 & 0.023 \\ \hline
\end{tabular}
\caption{The mean ($\mu_\be$), the standard deviation ($\sigma_\be$), the
skewness ($S_\be$) and the kurtosis ($K_\be$) for the scaled largest
eigenvalue data  from
 simulations in GOE ($N=500$), GUE ($N=100$) and GSE ($N=100$).  In each
 ensemble there were 5000 trials.}
\end{center}
\vspace*{-6mm}
\end{table}
\par
\bfigt\vspace*{-24mm}\hspace*{22mm}\epsfysize=125mm \epsffile{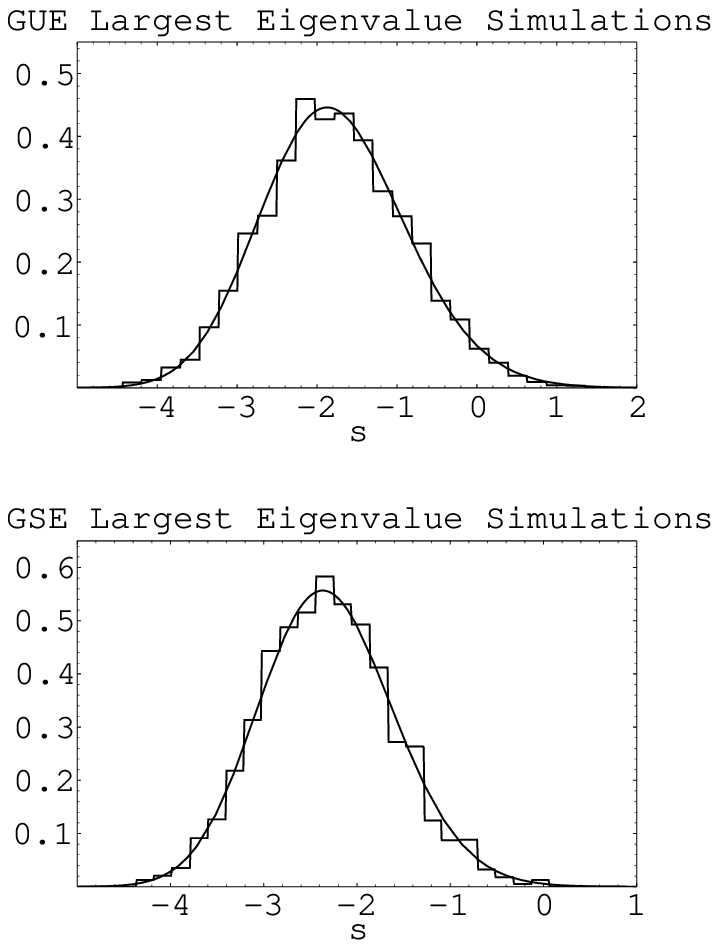}
\vspace*{-10mm}
\caption{Histograms for the scaled largest eigenvalues in GUE
and GSE each with $N=100$ and 5000 trials.  Solid curves are the limiting
densities $f_2(s)$ and $f_4(s)$, respectively.}
\vspace*{-5mm}
\efig
\par
\setcounter{equation}{0}\renewcommand{\theequation}{7.\arabic{equation}}
\noi{\bf 7.  Questions of Universality}
\par
Recall that the finite $N$ GOE is the
unique\footnote[2]{Uniqueness is up to centering and choice of standard
deviation on,
say, the off-diagonal matrix elements.}
 measure on the space of $N\times N$
real symmetric matrices $A$
that is both invariant under all orthogonal transformations and the matrix
elements $A_{jk}$
with $j\ge k$ are statistically independent.  Similar statements hold for both
the GUE and
GSE.

It is of interest to inquire to what extent  the various scaling limits of
these finite $N$
Gaussian ensembles are universal.  For the edge scaling limit,  the problem
is to characterize the domain of attraction of the limiting laws $F_\be$,
$\be=1,2,4$,
as given by (\ref{F1})--(\ref{F4}). The problem as so stated is open but there
has been recent
progress in showing that this domain of attraction
 contains more than the Gaussian case.   The most
studied case is that of  invariant random matrix models of Hermitian matrices
where
the measure is of the form
\[  \mu_N(dA) = Z_N^{-1} \exp\left(-\mbox{Tr}(V(A)\right) \, dA \]
and $V$ is a polynomial.  The Gaussian case corresponds, of course, to
$V(x)={1\ov 2} \be x^2$.

In \cite{KF} the cases $V(x)={1\ov 2} x^4$ and $V(x)={1\ov 12} x^6$ are
analyzed.
The nonuniveral part of the problem is the transformation   to
the scaling variable $s$. For example, in the quartic case (\ref{sVariable}) is
replaced by
\[ t= D_N + {s\ov (18 D_N^5)^{1/3}}, \>  D_N=2(N/12)^{1/4}. \]
But once this is done, it follows from the results in \cite{KF} that the
distribution
of the largest eigenvalue converges in the edge scaling limit, as defined
by the above equation, to the Gaussian result
$F_2(s)$.

In \cite{BI} there is  a
an account of the Hermitian matrix model with measure of the form
\[\mu_N(dA) = Z_N^{-1} \exp\left(-N\mbox{Tr}(V(A)\right) \, dA, \]
with
\[ V(x) = {t\ov 2} x^2 + {g\ov 4} x^4, \ \ t<0, \, g>0.\]
Note the insertion of the factor $N$ into the exponential.  This has the effect
that
the density of eigenvalues $\rho_N(z)$ converges without any rescaling of
variables to a limiting
density $\rho(z)$ given by
\[ \rho(z)= { g \vert z \vert \ov 2 \pi} \sqrt{(z^2-z_1^2)(z_2^2-z^2)}\]
with
\[ z_{1,2}= \left(-t\mp 2\sqrt{g}\ov g\right)^{1/2}. \]
  In this set up the edge of the spectrum remains bounded and the
transformation
 to the edge scaling variable $s$ is now
\[ z = z_2 + {s\ov c N^{2/3}},\> c=2^{1/3} g^{1/2}\, z_2.\]
It is proved \cite{BI} that the distribution  of the largest
eigenvalue converges to the Gaussian result $F_2(s)$.

Physicists (see \cite{brezin} and
references therein) have heuristic arguments that suggest that the domain of
attraction of
the  invariant matrix models for Hermitian matrices contains all those
potentials
$V$ for which the density  vanishes at the edge of the spectrum as does
the Wigner semi-circle, i.e.\ a square root.  By tuning the potential one
obtains
densities  that vanish faster than a square root---these models  will be in
a different universality class with regards to the edge scaling limit.

\begin{center}{\bf Acknowledgements}\end{center}
The first author wishes to thank Jan Felipe van Diejen and Luc Vinet for the
invitation
to speak
and the kind hospitality at the Workshop on Calogero-Moser-Sutherland Models.
This work was supported in part by
the National Science Foundation through grants DMS--9303413 (first author) and
DMS--9424292 (second author).
\par

 \end{document}